\documentclass[aps, prl, twocolumn, superscriptaddress, floatfix, nobibnotes]{revtex4-1}
\usepackage{float}
\usepackage{hyperref}
\usepackage{orcidlink}
\usepackage{blindtext}
\usepackage{xcolor}
\usepackage{amsmath, amssymb}
\usepackage{hyperref}
\usepackage{bm}
\usepackage{pdfpages}
\usepackage{etoolbox}

\hypersetup{colorlinks=true,
	linkcolor=blue,
	anchorcolor=blue,
	citecolor=blue,
	filecolor=blue,
	urlcolor=blue,
	bookmarksnumbered=true}

\makeatletter
\patchcmd{\@outputpage@head}{\@ifx{\LS@rot\@undefined}{}{\LS@rot}}{}{}{}
\makeatother

\newcommand{\meV}{~\mathrm{meV}}
\newcommand{\nm}{~\mathrm{nm}}
\usepackage{xspace}
\def\sc{{superconductor}\xspace}
\def\scing{{SC}\xspace}

\usepackage{physics}
\usepackage[subrefformat=parens,labelformat=parens,caption=false]{subfig}
\usepackage{stackengine}

\usepackage{xifthen}
\newcommand{\fref}[2][]{\ifthenelse{\isempty{#1}}{\hyperref[#2]{\ref*{#2}}}{\hyperref[#2]{\ref*{#2}(#1)}}}

\newcommand\maintitle{Anomalous Josephson effect in hybrid superconductor-hole systems}
\def\titlecommand{\maintitle}

\begin{document}
\title{\titlecommand}

\author{Peter D. Johannsen}
\affiliation{Department of Physics, University of Basel, Klingelbergstrasse 82, 4056 Basel, Switzerland}
\author{Henry F. Legg}
\affiliation{Department of Physics, University of Basel, Klingelbergstrasse 82, 4056 Basel, Switzerland}
\affiliation{SUPA, School of Physics and Astronomy, University of St Andrews,
	North Haugh, St Andrews, KY16 9SS, United Kingdom}
\author{Stefano Bosco}
\affiliation{Department of Physics, University of Basel, Klingelbergstrasse 82, 4056 Basel, Switzerland}
\affiliation{QuTech, Delft University of Technology, Delft, The Netherlands}
\author{Daniel Loss}
\affiliation{Department of Physics, University of Basel, Klingelbergstrasse 82, 4056 Basel, Switzerland}
\author{Jelena Klinovaja}
\affiliation{Department of Physics, University of Basel, Klingelbergstrasse 82, 4056 Basel, Switzerland}
\date{\today}

\begin{abstract}
We consider hybrid systems consisting of a hole-doped semiconductor coupled to electronic states of finite-size superconductors, where the opposite sign of the masses in the two subsystems give rise to insulating gaps at subband anticrossings. Consequently, increasing the coupling strength to the superconductor can paradoxically suppress the proximity-induced superconductivity in the semiconductor by enhancing these insulating gaps. We demonstrate that the presence of such induced insulating gaps leads to a characteristic anomalous behavior of the critical supercurrent in Josephson junctions based on these hybrid structures. Our findings provide important insights for the design of robust quantum computing platforms utilizing hybrid superconductor–hole systems.
\end{abstract}
\maketitle

\textit{Introduction.}
Over the past decade, the usage of hole gases in semiconductor devices --- such as germanium (Ge) --- has surged, in part due to the prospective scalability of these devices for future quantum computing purposes and in part because of the recent technological improvements in fabrication~\cite{Scappucci2020,Pillarisetty2011,Burkard2023}.
Furthermore, hole gases present strong tunable spin-orbit interactions, allowing for electrostatic control of system properties~\cite{Scappucci2020,Kloeffel2011,Hao2010,Hu2011,Terrazos2021,Froning2021,Liu2022,Adelsberger2022,Bulaev2005b,Bosco2021a,Bosco2021b,Bosco2022,Froning2021,Carballido2024,Abadillo2023,Hendrickx2024,Wang2024}, and tunable coupling to nuclear spins~\cite{bosco21,hu07,Fischer2008}.
Many potential quantum computing platforms utilize hole gases~\cite{Camenzind2022,Geyer2024,Liles2024,Wang2024b,Borsoi2023,Zhang2024,Jirovec2021,Saezmollejo2024}.
However, hybrid superconductor-semiconductor devices, in which thin superconducting (\scing) layers are brought into contact with a semiconductor such as a two-dimensional hole (electron) gas [2DHG (2DEG)]~\cite{Xiang2006,Valentini2024,ValentiniThesis,Chang2015}, not only host Andreev bound states --- the building block of Andreev spin qubits~\cite{Lee2014,Park2017,Hays2021,Spethmann2022} --- but they have also recently been proposed as platforms for topological superconductivity~\cite{Kitaev2001,Luethi2023a,Luethi2023b,Nayak2008,Maier2014,Laubscher2024}.

Central to hybrid superconductor-semiconductor systems is the superconducting proximity effect, in which a parent \sc induces a \scing pairing in a semiconductor.
The proximity effect in electron systems has been the subject of extensive study, both experimentally and theoretically~\cite{reeg17,reeg18a,reeg18b,Kopnin2011,vanLooInSb2023,SutoInAs2022,MayerInAs2019,TakayanagiInAs1985}.
While the proximity effect in hole gases exhibits many similarities to its electron counterpart, there is also a unique phenomenology and this can result in novel effects~\cite{Adelsberger2023,stanescu2022,Xiang2006,Valentini2024,ValentiniThesis,Chang2015,Hendrickx2018,Ridderbos2018,Hendrickx2019,Vigneau2019,Ridderbos2019,Aggarwal2021,Tosato2023,Zhuo2023,Moghaddam2014,babkin2024,MichelPino2025,Steele2024,Lakic2025}.
For instance, surprisingly, it was recently observed that a strong coupling between a hole gas in Ge and a \sc can reduce the proximity effect~\cite{Valentini2024,ValentiniThesis}, which would lead to undesirable device characteristics for quantum computing applications.

\begin{figure}[t!]
	\includegraphics[width=\linewidth]{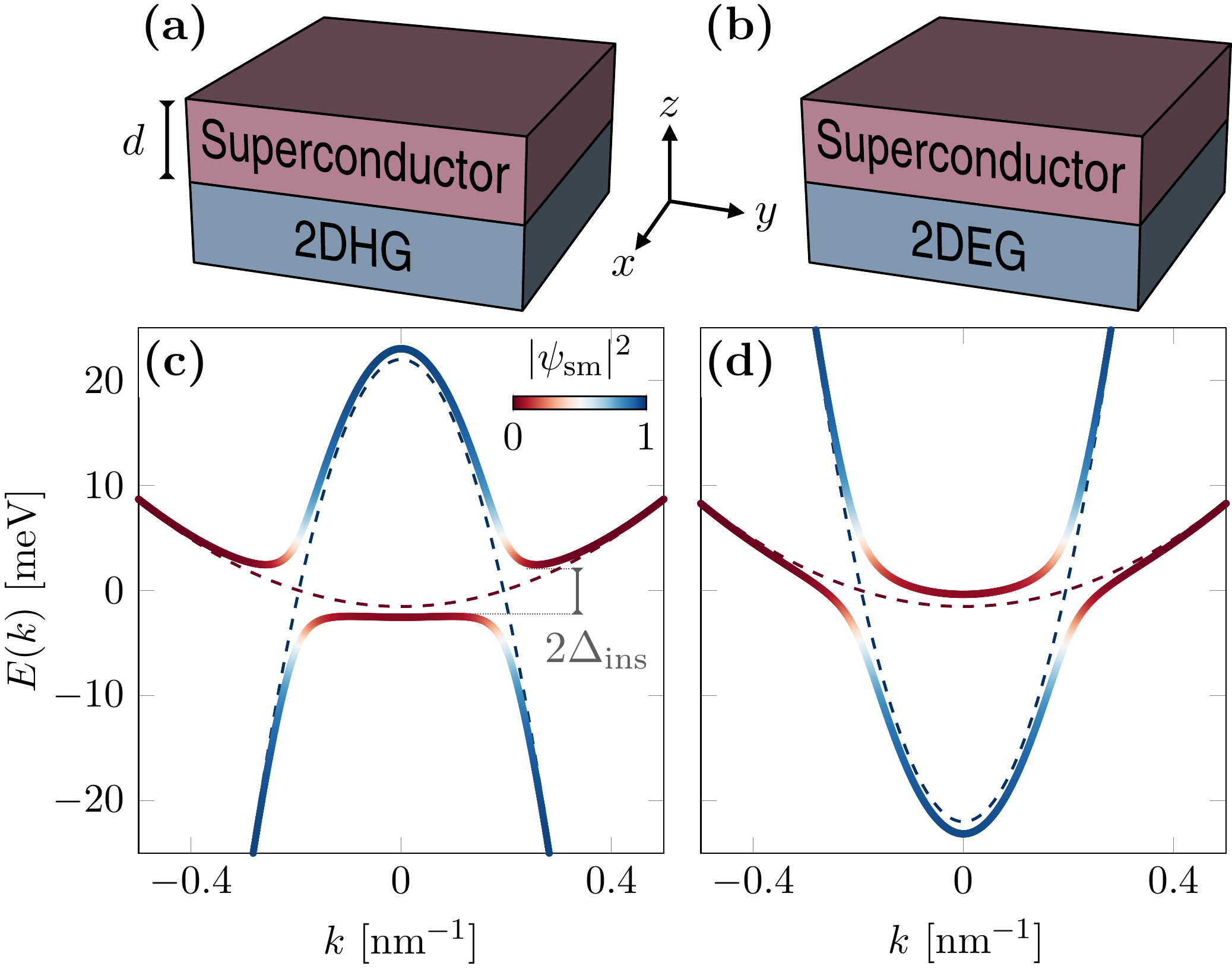}
	\caption{
		A hybrid superconductor-semiconductor system with a \scing layer of thickness, $d$, on top of a semiconductor [(a) 2DHG and (b) 2DEG].
		The tunnel coupling $t$ between a semiconductor and the \scing subband with smallest momentum can result in an insulating gap, $2\Delta_{\rm ins}$ [indicated by the gray bar in (c)], or an avoided crossing [as is shown in (d)],  depending on the sign of the mass.
		The hybridization between the subbands can be characterized by the weight in the semiconductor, $\vert\psi_{\rm sm}\vert^2$, which is indicated by the color of the dispersions in (c) and (d) [see colorbar in (c)].
		Here, $m_\mathrm{sc} = 0.95 m_e$ and $m_\mathrm{sm} = \pm 0.07m_e$, where $m_\mathrm{sc}$ and $m_\mathrm{sm}$ refer to the effective masses in the \sc and semiconductor, respectively~\cite{Luethi2023a,Luethi2023b,Marcelina2017}, with $m_e$ the bare electron mass.
		The parameters used are $t = 5\meV$, $\mu_{\rm sm} = 22\meV$, $\Delta_0=0$, and the effective SC subband chemical potential is ${\bar \mu}_{\rm sc}=1.5\meV $.
	}\label{fig1}
\end{figure}
In hybrid devices, the regime of most interest is when the proximity-induced pairing is of the same order of magnitude as in the parent \sc. If the crossing between  a subband of a 2DEG and  a metallic subband in the \sc is close to the chemical potential $\mu$ of the coupled system, the proximity effect is  most pronounced [see Fig.~\fref[d]{fig1}].
In this case, the strong hybridization between the two subsystems results in a substantial \scing gap in the semiconducting subsystem~\cite{reeg17,reeg18a,reeg18b}.
In contrast, due to the opposite signs of the effective masses, coupling a 2DHG subband to a metallic subband results in an \textit{insulating} gap in the energy spectrum [see Fig.~\fref[c]{fig1}], such that there are no states with a significant weight in the semiconductor in this energy range.
The absence of states at the chemical potential in the 2DHG suggests that the \scing proximity effect will be suppressed close to such subband crossings. As a result, in such a setup, we expect an interesting interplay between these insulating and \scing gaps.

Interestingly, we find that the presence of insulating gaps can imply that \textit{increasing} the coupling between the subsystems can result in a \textit{reduction} of the proximity-induced \scing pairing in the semiconductor.
We demonstrate how this effect can be observed in one-dimensional (1D) Josephson junctions (JJs).
Our results are most relevant to systems in which \scing subbands are well-defined, \textit{i.e.} the mean-free path in the \sc is longer than the layer thickness, which in turn is assumed to be much shorter than the \scing coherence length.
This is a regime only recently accessible experimentally and could explain the inconsistent \scing proximity effect recently observed in hybrid Ge devices~\cite{Valentini2024,ValentiniThesis}.
Our main results are valid both for one- and two-dimensional coupled systems~\cite{reeg17, reeg18a, reeg18b}, whereas the JJ analysis is done for nanowire systems.


	{\it Model.} To model the proximity effect in hybrid \sc-semiconductor devices, we consider a thin 2DHG (2DEG) coupled to a \scing layer of thickness $d$ (along $z$ direction, see Fig.~\fref{fig1}).
The \sc is assumed to have a parabolic dispersion with positive mass $m_\mathrm{sc}$, an $s$-wave pairing amplitude $\Delta_0$, 
and to be
described by the Bogoliubov-de-Gennes (BdG) Hamiltonian
\begin{equation}
	H_{\mathrm{sc},k} = \left(\frac{k^2-\partial_z^2}{2m_\mathrm{sc}} - \mu_{\rm sc}\right)\eta_z + \Delta_0\eta_x, \label{eq:Hsc}
\end{equation}
where $\eta_{i}$ denotes the $i$-th Pauli-matrix in Nambu space,  $\mu_{\mathrm{sc}}$ is the chemical potential in the \sc 
and  $k = (k_x^2+k_y^2)^{1/2}$ is the in-plane momentum  ($\hbar = 1$).
 The semiconducting layer consists of a single spin-degenerate parabolic subband with a negative (positive) effective mass $m_{\mathrm{sm}}$, describing a 2DHG (2DEG) subband, such that the Hamiltonian is given by
\begin{equation}
	H_{\mathrm{sm},k} = \left(\frac{k^2}{2m_{\mathrm{sm}}} - \mathrm{sgn}(m_{\rm sm})\mu_{\rm sm}\right)\eta_z, \label{eq:Hsm}
\end{equation}
where $\mu_{\mathrm{sm}}$ is the chemical potential in the semiconductor.
We use $\mathrm{sgn}(m_{\rm sm})\mu_{\rm sm}$ to ensure that $\mu_{\rm sm}>0$ gives rise to occupied states in the semiconductor. We note that, due to the rotational-invariance of the Hamiltonians defined in Eqs.~\eqref{eq:Hsc} and~\eqref{eq:Hsm}, the energy spectra depend only on $k$.
Thus, the results in this work also pertain to 1D systems, see Refs.~[\citenum{reeg17}] and~[\citenum{reeg18a}].

We are interested in the low-energy physics in the subband of the semiconductor, which, especially in the presence of strain and in the absence of inversion-symmetry breaking, implies that it is appropriate to use the effective mass approximation~\cite{Terrazos2021}.
We note that the full Luttinger-Kohn description does not modify the results in this work.
The coupling between the \sc and semiconductor is assumed to be independent of momentum and to occur at the interface $z=z_0$~\cite{reeg18a}.

In this work we focus on the parameter regime where there is only a single \scing subband that is strongly tunnel coupled to the semiconductor subband (with amplitude $t$).
Therefore, we only need to consider the metallic subband with the smallest momentum close to $\mu$.
All other subbands, although present, are assumed not to contribute to the effects of interest, and are therefore not shown in Fig.~\fref[a,b]{fig2}.
The subband spacing in the \sc due to the hard-wall confinement, $\delta E_{\rm sc} = \pi v_{\rm F}/d \approx 75\meV$, and $\mu_{\rm sc}$ are assumed to be the largest energies in the system ($v_{\rm F}$ is the Fermi velocity in the \sc).

	{\it Proximity effect.} The proximity effect can be taken into account using the self-energy, $\Sigma_{k,\omega}$, obtained by solving the Dyson equation for the full Matsubara Green function of the semiconductor
\begin{equation}
	G_{k,\omega} = \left( i\omega - H_{\mathrm{sm},k} - \Sigma_{k,\omega}\right)^{-1},
\end{equation}
where $\omega$ is a  Matsubara frequency.
In particular, the self-energy can be written as~\cite{reeg17, reeg18a}
\begin{equation}
	\Sigma_{k,\omega} = (i\omega -\Delta_0 \eta_x) (1 - \Gamma_{k,\omega}^{-1})-\delta\mu_{k,\omega}\eta_z\label{eq:Sigma}.
\end{equation}
Here,
$(1 - \Gamma_{k,\omega}^{-1})$ and $\delta \mu_{k,\omega}$ have simple poles at the eigenvalues of the \sc.
The self-energy is characterized by the coupling strength, $\gamma = \rho_{\rm 1D} t_0^2$, where $t_0$ is the coupling amplitude, and $\rho_{\rm 1D}$ is the 1D density of states of a \sc at the chemical potential in the normal state. We emphasize that $\Sigma_{k,\omega}$ does not depend on $H_{{\rm sm},k}$ and thus describes the self-energy due to the \sc for both $d$DEG and $d$DHG, for $d=1,2$.

\begin{figure}[t!]
	\includegraphics[width = \linewidth]{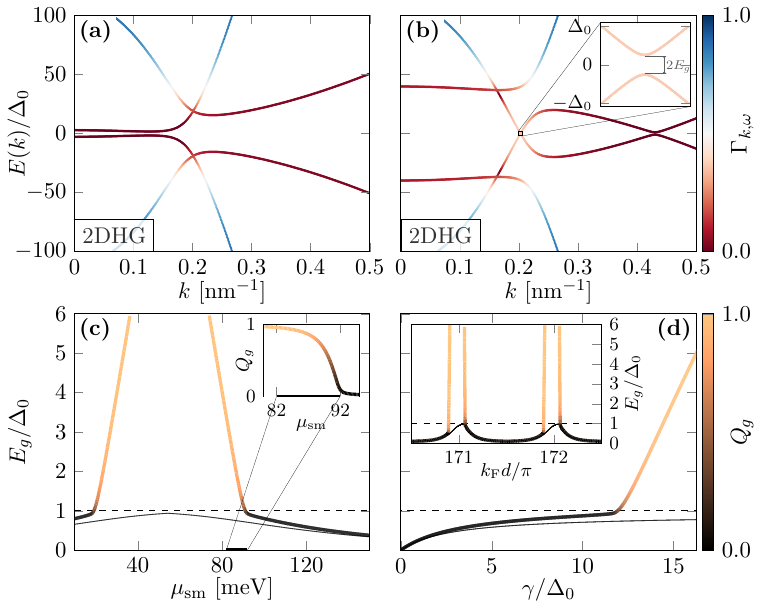}
	\caption{
		The induced gap $E_g$ in the spectrum $E(k)$ of a hybrid \sc-2DHG  can be (a) larger  or (b) smaller  than $\Delta_0$ of the \sc.
		While for proximity-induced pairing gap  $E_g \leq \Delta_0$  [see the insert in (b)], in the insulating phase (a) we find $E_g > \Delta_0$.
		The insulating phase emerges when the 2DHG and \scing subbands cross close to $\mu$ [see also Fig.~\fref[c]{fig1}].
		The relative position of the subbands can be tuned by adjusting (c) $\mu_{\rm sm}$ or inset (d) $k_{\rm F} d$.
		For 2DEGs [thin lines in (c,d)], $E_g \leq \Delta_0$  corresponds to the \scing proximity gap. In contrast, 2DHGs are in the insulating phase at the same parameter values  at which  the \scing pairing is at  its maximum   in the 2DEG.
		When $E_g > \Delta_0$,  the BCS charge, $Q_{g}$,  approaches unity, implying a suppression of the proximity pairing. In contrast, $Q_{g} \approx 0$ when $E_g \leq \Delta_0$,   implying an almost equal linear combination of quasiparticles and quasiholes, and thus $Q_g$ is a good measure for the breaking-down of pairing.
		(d) As expected for a 2DEG (thin line),
		$E_g \to \Delta_0$ as $\gamma \gg \Delta_0  $. 
		In contrast, for a 2DHG (thick line) $E_g \gg \Delta_0$ for large $\gamma$ with $Q_g \to 1$,
		i.e. increasing $\gamma$ strongly \textit{reduces} the \scing proximity effect.
		In (a,b) $\gamma = \Delta_0$, (c) $\gamma = 2\Delta_0$, and inset (d) $\gamma = 25\Delta_0$, where $\Delta_0 = 0.2\meV$, 
		while $k_{\rm F}d/\pi=172, 172.05, 172.025, 171.05$ in (a,b,c,d), resp.
		We set $d=30\nm$ and vary $k_{\rm F}$, additionally $\mu_{\rm sm} = 22\meV$ in (a,b,d), and $\mu_{\rm sc} = 13~\mathrm{eV}$ everywhere.
		Also, $z_0= \lambda_F/7$, where $\lambda_F \approx 0.35\nm$ is the Fermi wavelength in the \sc.
		The effective masses are as in Fig.~\ref{fig1}.
	}\label{fig2}
\end{figure}

\textit{Renormalized dispersions.} After analytic continuation, we numerically determine the poles of the full retarded Green function,  $G^R_{k,\omega}$, which define the renormalized dispersion relation  $E(k)$  in the semiconductor. The renormalized energy spectrum is shown in Fig.~\ref{fig2}, comparing (a) the insulating and (b) the proximitized \scing phases.
We define the gap in the energy spectrum as $E_g = \min_{k < k_c} |E(k)|$, where the cutoff $k_c$  is chosen to capture the minimum of the semiconductor band at the smallest momentum. Additionally, we define $k_g$ as the momentum where $E(k_g) = E_g$, and $(H_{{\rm sm},k_g} + \Sigma_{k_g,E_g})\ket{k_g,E_g} = E_g \ket{k_g,E_g}$. Importantly, $E_g$  corresponds to the excitation gap in the semiconductor, which does not necessarily match the proximity-induced \scing pairing gap.
Remarkably,  $E_g$  can exceed  $\Delta_0$  if the system is in the insulating phase, where an intrinsic insulating gap $2\Delta_{\rm ins}$ exists even when  $\Delta_0 = 0$  [see Fig.~\fref[c]{fig1}].
The insulating gap in the normal system is given by
\begin{align}
 	\Delta_{\rm ins} = \frac{2t\sqrt{ m_{\rm sc} \vert m_{\rm sm}\vert }}{  m_{\rm sc} + \vert m_{\rm sm}\vert   } = \frac{2t\sqrt{ v_{\rm sc} \vert v_{\rm sm}\vert }}{v_{\rm sc} + \vert v_{\rm sm}\vert },
\end{align}
which can exceed $\Delta_0$. Here $v_{\rm sc}$ ($v_{\rm sm}$) is the Fermi-velocity of the relevant superconducting (semiconducting) subbands.

This is indeed possible when the chemical potential is close to the anticrossing between the 2DHG and metallic subbands.
Moreover,  $E_g$  for 2DHGs can exceed the corresponding value for 2DEGs even when  $E_g < \Delta_0$  [see Fig.~\fref[c,d]{fig2}], without significantly suppressing \scing properties.
This suggests that the insulating gap can enhance the proximity gap while maintaining robust pairing.

In this work, we define the insulating phase as $E_g > \Delta_0$, however, it should be noted that the transition between \scing and insulating phases is gradual, especially on the scale of the temperature of \scing systems, and on the scale of $\Delta_0$, as can be seen in Fig.~\fref[c]{fig2}.
However, for the purpose of this work, we will treat the phases as distinct.

As shown in Fig.~\fref[c,d]{fig2}, insulating phases
and \scing phases with strong proximity-induced \scing pairing are close to each other.
As such, extra care is required to differentiate between a semiconductor with a large pairing gap and one that is insulating, which is difficult to determine from the density of states alone or from the spectrum.
However, when a supercurrent is forced through the semiconductor, as in a JJ, we can expect distinct signatures of insulating and \scing phases, see below.

The resonances in the insert of Fig.~\fref[d]{fig2}, 
where the subband crossing occurs exactly at $\mu$, arise approximately when 
$k_{\rm F} d / \pi$ is an integer [see Fig.~\fref[a]{fig2}]. 
The insulating gap persists in an interval 
$\sim2\gamma/\delta E_{\rm sc}$  around $k_{\rm F} d / \pi$ (assuming $\Delta_0,\gamma \ll \delta E_{\rm sc}$).
As such, increasing the coupling between the systems and/or the \scing layer thickness will not only improve the proximity effect away from resonance (as discussed in~\cite{reeg18a}
for 2DEGs) but will also make it more likely to be in an insulating phase.
Numerically, we find that this behavior persists until $\gamma \gtrsim \delta E_\mathrm{sc}$, at which point the insulating phase starts to disappear and we obtain similar behavior as in
Ref.~\cite{reeg18a} such that the \scing proximity effect is more accurately described by the infinite \sc limit, and there is no difference in the proximity effect between 2DHGs and 2DEGs.

Surprisingly, by increasing $\gamma$ a system being initially in the \scing phase can be brought into the insulating phase for chemical potentials close to the crossing point, see 
Fig.~\fref[d]{fig2}.
First, $E_g$ increases as a function of $\gamma$ for the 2DHG (thick line), so does the BCS charge $Q_g= \vert \langle k_g, E_g\vert \eta_z\vert k_g, E_g\rangle\vert$  (line color).
The imbalance of the quasiparticle and quasihole components of the wavefunction leads to  $Q_g \approx 1$, indicating a reduction of the \scing proximity effect. Conversely, in the 2DEG case, $E_g$ always represents the \scing gap and thus approaches $\Delta_0$ as $\gamma/\Delta_0 \gg 1$, with $Q_g\approx 0$ everywhere [thin line in 
Fig.~\fref[d]{fig2}].

As discussed above, in this work we focus on the intermediate coupling regime, where $\Delta_0 \lesssim \gamma\ll\delta E_{\mathrm{sc}}$, such that we can obtain a large proximity effect even away from resonance, but still only need to consider an individual subband in the \sc.
We note that this is the experimentally relevant regime because the \scing pairing is usually $\Delta_0\lesssim 1$~meV, while the subband spacing in metallic layers has a range $\delta E_{\mathrm{sc}}\sim 10 - 100$~meV~\citep{Tosato2023}. It is important to note that in this regime one also expects the largest differences between 2DHG and 2DEG hybrid systems with the insulating gap dominating over the \scing proximity gap.
As shown in Figs.~\fref[c-d]{fig2}, whether the coupling between the subsystems leads to a \scing or insulating gap can strongly depend on the system parameters.
In an experiment, it is impractical to vary $d$ and $k_{\rm F}$ will be almost fixed.
Therefore, upon fabricating a device, it is possible to end up close to a resonance and, with only a small variation of parameters, either obtain a large proximity-induced pairing or an insulating phase. 
Here we treat the effective parameters such as $\gamma$, $d$, $z_0$, $\mu_{\rm sm}$ as independent. However, for example, applying an electric field to the semiconductor can localize the semiconductor wavefunction closer to the interface to the \sc and this
can increase $\gamma$, change $\mu_{\rm sm}$ as well as influence the effective lengths $d$ and $z_0$.

\begin{figure}
	\centering
	\includegraphics[width=\linewidth]{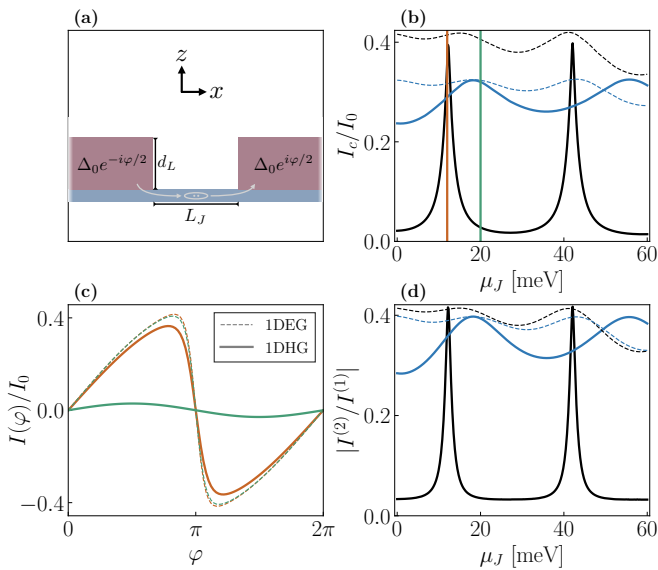}
	\caption{
	         1D JJ consisting of an infinite 1D semiconducting wire (blue), with a single subband that extends through both the junction region and the leads.
		Each lead is covered by a semi-infinite \scing film (brown) with a phase difference $\varphi$ across the JJ that generates a supercurrent $I(\varphi)$ in the wire.
		When the leads are in the insulating regime (black lines), the critical current $I_c$ [panel (b)] and the ratio between the first two harmonics of the CPR, $\vert I^{(2)}/I^{(1)}\vert$, [panel (d)] are strongly non-monotonic:
		Adjusting the chemical potential $\mu_J$ in the junction reveals sharp peaks in $I_c$ and $\vert I^{(2)}/I^{(1)}\vert$ [see (b,d)], corresponding to bound states confined around the normal junction region.
		In contrast, for a 1DEG coupled to a \sc [dashed lines in (b,d)] or when the 1DHG is in the \scing phase [blue lines in (b,d)], the system exhibits a more steady dependence of $I_c$ and $\vert I^{(2)}/I^{(1)} \vert$ on $\mu_J$.
		In these cases, the oscillations are due to Andreev bound states whose transmissivity is $\mu_J$-dependent.
		The vertical lines in (b) show $\mu_{J}$ for the CPRs $I(\varphi)$ in (c) [$\mu_J = 12\meV$ (orange) and $20\meV$ (green)].
		The junction length is $L_J = 50~\rm nm$, the \sc thickness  $d_L = 20~\rm nm$, and the lattice constant $a=0.5\nm$.
		Here, $k_BT = \Delta_0/20\approx 100~\rm mK$, $\mu_{\rm sm} = 90\meV$, and the hopping between the semiconductor and superconductor $t_{\rm c} = 50\meV$.
		For the black lines in (b,d) and all CPRs in (c) $\mu_{\rm sc} \approx 293~\mathrm{meV}$, while for the blue lines in (b,d) $\mu_{\rm sc} \approx 306~\mathrm{meV}$.		The effective masses are as in Fig.~\ref{fig1}. 
		The supercurrents are normalized relative to $I_0 = \vert e\vert\Delta_0 \approx 49~\rm nA$.
	}\label{fig3}
\end{figure}

{\it Anomalous Josephson effect.}
We consider a 1D JJ, where the leads consist of a semiconductor (one subband) covered by a superconductor with finite width, while the junction region consists only of this semiconducting subband. A schematic of the setup is shown in Fig.~\fref[a]{fig3}. The system is modelled using a tight-binding Hamiltonian, which consists of a junction region (in which only semiconducting sites are present), and of two infinite leads. Following Refs.~[\citenum{Hess2023,Yeyati1995,deVries2018,Himmler2022}] and using \href{https://kwant-project.org}{KWANT}~\cite{kwant} we numerically evaluate the 
supercurrent $I(\varphi)$ as function of the phase difference $\varphi$ across the JJ.

As discussed above, the leads can be in the insulating phase for the semiconducting subband, which could suppress proximity-induced superconductivity.
Consequently, the critical supercurrent $I_c$ is strongly suppressed since other high-momentum superconducting subbands have a relatively small weight in the semiconductor such that they do not couple strongly to the normal semiconducting region.
To illustrate this suppression, we plot $I_c$ as a function of the chemical potential $\mu_J$ in the junction both in- and outside the insulating regime for the leads, see Fig.~\fref[b]{fig3}.
As expected, near the insulating phase [Fig.~\fref[b]{fig3}], $I_c$ for 1DHGs (full line) is significantly suppressed compared to 1DEGs (dashed line).
Notably, in this parameter regime, the hole system only exhibits a substantial $I_c$ when a bound state forms in the normal region at an energy within the parent \scing gap $\Delta_0$, leading to a highly anomalous Josephson effect.
This is evident from the current peaks observed in Fig.~\fref[b]{fig3}.
We note that the sharp peaks in $I_c$ versus $\mu_J$ remain, even when the standard deviation of the chemical potentials (due to disorder) exceeds $\Delta_0$.

Similarly, Fig.~\fref[d]{fig3} shows the ratio of the first two Fourier components of the current-phase relation (CPR) $I (\varphi)$, with $\varphi$ the superconducting phase difference across the JJ, 
both in- and outside the insulating phase.
These results further highlight a strong suppression of transparency in the insulating phase~\cite{Ciaccia2024,Willsch2024,Legg2023}.
Finally, away from the insulating regime [blue lines in Figs.~\fref[b]{fig3} and~\fref[d]{fig3}], the anomalous Josephson effect disappears, and both 1DHG (solid) and 1DEG (dashed) systems exhibit similar critical current and transparency behavior.

	We conclude by noting that the effects  discussed here are robust against moderate disorder, e.g., variations in chemical potential.
	As such, the anomalous behavior uncovered in this work is not only of fundamental importance but also of key relevance in hybrid systems, such as proximitized Ge hole semiconductors, which are of significant current experimental interest~\cite{Hendrickx2018,Ridderbos2018,Hendrickx2019,Vigneau2019,Ridderbos2019,Aggarwal2021,Tosato2023,Zhuo2023,Valentini2024,ValentiniThesis}.

{\it Acknowledgements.}
We would like to thank Joel Hutchinson, Maximilian Hünenberger, and Christoph Adelsberger for useful discussions.
This work is supported by the Swiss National Science Foundation (SNSF) and NCCR SPIN (Grant No. 51NF40-180604).

%

\end{document}